\documentclass[12pt]{iopart}
\usepackage[dvips]{graphicx}
\usepackage{iopams}
\usepackage[dvips]{epsfig}
\usepackage{figlabel}

\newcommand{\hs}[1]{\hspace*{#1cm}}
\newcommand{\vs}[1]{\vspace*{#1cm}}
\newcommand{\la}{\langle}
\newcommand{\ra}{\rangle}
\newcommand{\half}{{\textstyle \frac{1}{2}}}

\begin{document}

\title{The Fermionic Particle Density of Flat \boldmath{1+1} Dimensional Spacetime seen by an Arbitrarily Moving Observer.}
  
\author{Carl E. Dolby} \address{Department of Theoretical Physics, 1 Keble Road, 
Oxford, OX1 3RH, UK} 
\author{Mark D. Goodsell} \address{Hertford College, Catte Street, Oxford, OX1 3BW, UK}
\author{Stephen F. Gull} \address{Astrophysics Group, Cavendish 
Laboratory, Madingley Road, Cambridge, CB3 0HE, UK}

\begin{abstract}
A coordinate system is constructed for a general accelerating observer in 1+1 dimensions, and is used to determine the particle density of the massless Dirac vacuum for that observer. Equations are obtained for the spatial distribution and frequency distribution of massless fermions seen by this observer, in terms of the rapidity function of the observer's worldline. Examples that are considered include the uniformly accelerating observer as a limiting case, but do not involve particle horizons. Only the low frequency limit depends on the possible presence of particle horizons. The rest of the spectrum is `almost thermal' whenever the observer's acceleration is `almost uniform'.
\end{abstract}

\section{Introduction}
Unruh \cite{Un} and Davies\cite{Davies} seperately observed in th 1970's that an observer who accelerated uniformly 
through the Minkowski vacuum would see a ``thermal" spectrum of scalar particles, 
whose perceived temperature depends on the acceleration. The same result 
holds for fermions \cite{Soffel,Hughes,CaDe,IyKu,Hacyan,Takagi}. The constant 
acceleration case has been extensively investigated and reviewed \cite{Takagi,BD}. Most approaches to the constant acceleration case employ 
Rindler coordinates (\cite{Takagi,Horibe} are partial exceptions), relying on 
the symmetry properties of Rindler space, and often emphasizing the importance of 
the horizon. If the Killing symmetry of Rindler space or its associated horizon 
were indeed responsible for the result, then questions arise such as: would an 
observer who accelerates `almost constantly' see a spectrum of particles which 
is `almost thermal'? If an observer's acceleration is constant for a long but finite 
period of time, will that observer see an `almost thermal' spectrum during that 
period? If not, the stability of the Unruh result could be in question.
	
Evidence from detector models \cite{SrPa,Sanchez81A} supports the stability of the Unruh result for scalar fields. Sriramkumar \cite{SrPa} demonstrates that a detector which accelerates constantly, but which is turned on only for a finite time, will detect a spectrum of particles differing from the thermal result only by transients.

Bogoliubov approaches to scalar particle creation have been considered in some classes of accelerating frames in 1+1 dimensions \cite{Sanchez79,Sanchez81B,Yang} and in 3+1  dimensions \cite{SaWh}. The method of Yang et. al. \cite{Yang} was based on the `instantaneous rest frame' of a given observer. Interesting results were obtained, but their interpretation is limited by the multivalued nature of the instantaneous rest frame \cite{Dolbyradar,Misner2}. The `instantaneous rest frame' will in general assign more than one time to any event further than $1/g(t) $ from an accelerating observer -- it does not {\it foliate} the spacetime. This leads in \cite{Yang} to multivalued predictions for the particle content at any such event, raising problems of interpretation. The possible foliations considered in Sanchez \cite{Sanchez79,Sanchez81B} and Whiting \cite{SaWh} are similar to those used here, and will be discussed further in the next section.

Below we consider an arbitrarily moving observer in 1+1 dimensions, traveling through the massless fermionic Minkowski vacuum. The foliation we shall use is the ``radar time" construction described in full elsewhere \cite{Dolby2,Dolbyradar,Dolby} and presented briefly in Section 2. In Section 3 this construction is used to derive an expression for the number of fermions measured by this observer in the inertial vacuum, and also the spatial distribution of those fermions. Examples are presented in Section 4, showing that observers who accelerate `almost constantly' see an `almost thermal' spectrum in the causal envelope of their period of acceleration, and further supporting the stability of the Unruh result.

We use ``natural" units throughout such that $\hbar = c = 1 $, and restrict our study to $1 + 1 $ dimensions.

\section{Radar Coordinates}

The concept of radar coordinates for an accelerated observer is not new \cite{Dolbyradar,Pauri}. They provide a set of coordinates having one-to-one correspondence with the entire region of the Lorentz frame with which the accelerated observer can exchange signals. Consider an observer whose worldline $\lambda $ in an inertial frame is described by:

\begin{equation} x^{\pm} \equiv t \pm x = x^{\pm}_{\lambda}(\tau_{\lambda}) = \int^{\tau_{\lambda}} {\rm d} \tau \ e^{\pm \alpha(\tau)} \label{eq:1} \end{equation} 
where $\tau_{\lambda}$ is the observer's proper time, and 
$\alpha(\tau_{\lambda})$ is the observer's `rapidity' at time $\tau_{\lambda}$. 
The exponential $e^{\alpha(\tau_{\lambda})}$ is the obvious time-dependent generalization 
of the $k$ of Bondi's $k$-calculus \cite{Bondi,Bohm,Dinverno}.  The observer's 
acceleration is $a(\tau_{\lambda}) = d \alpha/d \tau_{\lambda}$, and satisfies $a^2 = - A^{\mu} A_{\mu}$. The 
observer's worldline is completely specified by the choice of origin 
(i.e. $x^{\mu}(0)$) and the rapidity function $\alpha(\tau_{\lambda})$, or 
by the choice of origin, the initial velocity, and the function 
$a(\tau_\lambda)$. The radar time $\tau(x^\mu)$ and radar distance 
$\rho(x^\mu)$ are defined by:

\begin{eqnarray}\tau \equiv \frac{1}{2}(\tau^+ + \tau^-) \label{eq:2} \\
\label{eq:rt}\rho \equiv \frac{1}{2}(\tau^+ - \tau^-) \nonumber \end{eqnarray}
where 

\begin{quote}
\begin{tabular}{lp{12cm}} $\tau^+ \equiv $ & time at which a signal from an event travelling in the observer's backward direction (i.e. $t + z = \textrm{constant}$) is received or sent.\\
$\tau^- \equiv $ & time at which a signal travelling in the observer's forward direction (i.e. $t - z = \textrm{constant}$) is received or sent. \end{tabular}
\end{quote}
and are such that $\rho$ is positive in the observer's forward direction and negative behind -- this choice is convenient for $1 + 1$ dimensions. From equations (\ref{eq:1}) and (\ref{eq:2}) it follows that $\tau^{\pm}$  are related to $x^{\pm}$ by 
$$ x^{\pm} = x^{\pm}_{\lambda}(\tau^{\pm}) $$
To see the connection between the $\tau$, $\rho$ coordinates used here, and the $t'$, $x'$ used in the scalar field work of Sanchez and Whiting \cite{Sanchez79,Sanchez81B,SaWh}, consider the most general transformation, written as equation (6.2) of \cite{Sanchez81B} and described briefly in \cite{SaWh}. This is of the form:

\begin{equation}  x^+ = g(x^{'+}) \mbox{ and } x^- = - f(-x^{'-}) \label{sanch1} \end{equation} 
where $f$ and $g$ are analytic functions and $x^{' \pm} = t' \pm x'$ are the `accelerated coordinates'. The coordinate $t'$ in general has no connection with the proper time of any particular observer. In radar coordinates, the line $\rho=0$ is the worldline of the observer. The requirement that $\tau$ coincide with the observer's proper time on this worldine gives:
\begin{equation} \frac{d x^+_{\lambda}}{d \tau} \frac{d x^-_{\lambda}}{d \tau} = 1 \label{proptime} \end{equation}

 It is this condition that uniquely assigns the choice of foliation to the choice of observer, and which allows $x^{\pm}_{\lambda}(\tau)$ to be written in terms of the observer's rapidity function, as in equation (1). Throughout \cite{Sanchez79}, and in the main body of \cite{Sanchez81B,SaWh}, attention is restricted to the case $f = g$. Combined with the restriction (\ref{proptime}), this leads to the time-reversal invariant case considered at the end of Section 3. 

	The metric in radar coordinates is:
$${\rm d} s^2 = e^{(\alpha(\tau^+) - \alpha(\tau^-))} ({\rm d} \tau^2 - {\rm d} \rho^2) $$ 

The `time-translation vector field' \cite{Dolby2,Dolby} is simply $k^{\mu} 
\frac{\partial }{\partial x^{\mu} } =  \frac{\partial }{\partial \tau}$, 
and the hypersurfaces $\Sigma_{\tau}$ are hypersurfaces of 
constant $\tau$. To illustrate these definitions we consider 
some examples.


\subsection{Inertial Observer}
\label{Inertial}

As a consistency check, consider an inertial observer with a 
velocity $v $ relative to our original frame. Then $\alpha$ is constant, 
and $x^{\pm}_{\lambda}(\tau_{\lambda}) = e^{\pm \alpha} \tau_{\lambda}$. The 
coordinates $\tau^{\pm}$ are now given by $\tau^{\pm} = e^{\mp \alpha} x^{\pm}$, so that 
$$\tau = \half(e^{- \alpha} x^+ + e^{\alpha} x^-) = \frac{t - v x}{\sqrt{1 - v^2}}, 
\hs{1} \rho = \half(e^{-\alpha} x^+ - e^{\alpha} x^-) = \frac{x - v t}{\sqrt{1 - v^2}}$$
   The radar coordinates of an inertial observer are just the coordinates of 
that observer's rest frame, as expected.

\subsection{Constant Acceleration}
\label{constant}

The simplest nontrivial case is that of constant acceleration \cite{Misner2}. In this case 
$\alpha(\tau) = a \tau$ and we have 
$x^{\pm}_{\lambda}(\tau_{\lambda}) = \pm a^{-1} e^{\pm a \tau_{\lambda}}$, which gives:
\begin{eqnarray} \tau & = \frac{1}{2 a} \log\left(\frac{x+t}{x-t}\right) \hs{2} \rho = \frac{1}{2 a} \log\left(a^2(x^2 - t^2)\right) \\
 {\rm d} s^2 & = e^{2 a \rho} ({\rm d} \tau^2 - {\rm d} \rho^2) \end{eqnarray}

\begin{figure}[h]
\vspace{-.3cm}
\center{\epsfig{figure=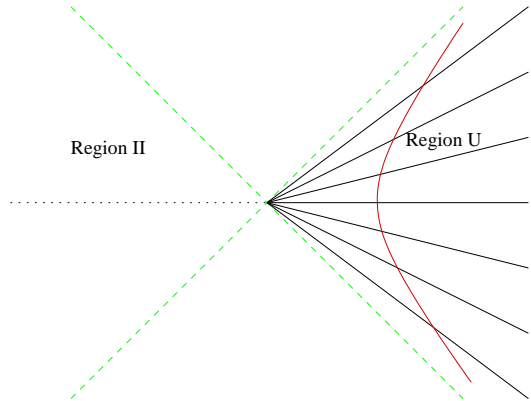 , width=7cm}}
\caption{{\footnotesize Hypersurfaces of simultaneity of a uniformly accelerating observer.}}
\vspace{-.3cm}
\end{figure}

These are Rindler coordinates, which cover only region U of Figure 1, as expected \cite{Dolbyradar,Pauri}.
The hypersurfaces of constant $\tau$ are given by $t_{\tau_0}(x) = x \tanh(a \tau_0)$.

\subsection{The `Smooth Turnaround' Observer}

Consider an observer (Alice say) with $\alpha(\tau_{\lambda}) = a \tau_c \tanh(\tau/\tau_c)$. Her trajectory is given by:

$$x^{\pm}_{\lambda}(\tau_{\lambda}) = \int_{0}^{\tau_{\lambda}} {\rm d} \tau \ e^{\pm a \tau_c \tanh(\tau/\tau_c)} $$
She has acceleration $a(\tau) = a \cosh^{-2}(\tau/\tau_c)$, so is uniformly accelerating for $|\tau| << \tau_c$ and is inertial for $|\tau| >> \tau_c$. As the parameter $\tau_c \rightarrow \infty$ this case approaches the constant-acceleration case. However, there are no horizons for this `smooth-turnaround' observer.

\begin{figure}[htb]
\figstep
\hs{1}\begin{minipage}[b]{6.5cm}
\epsfig{figure=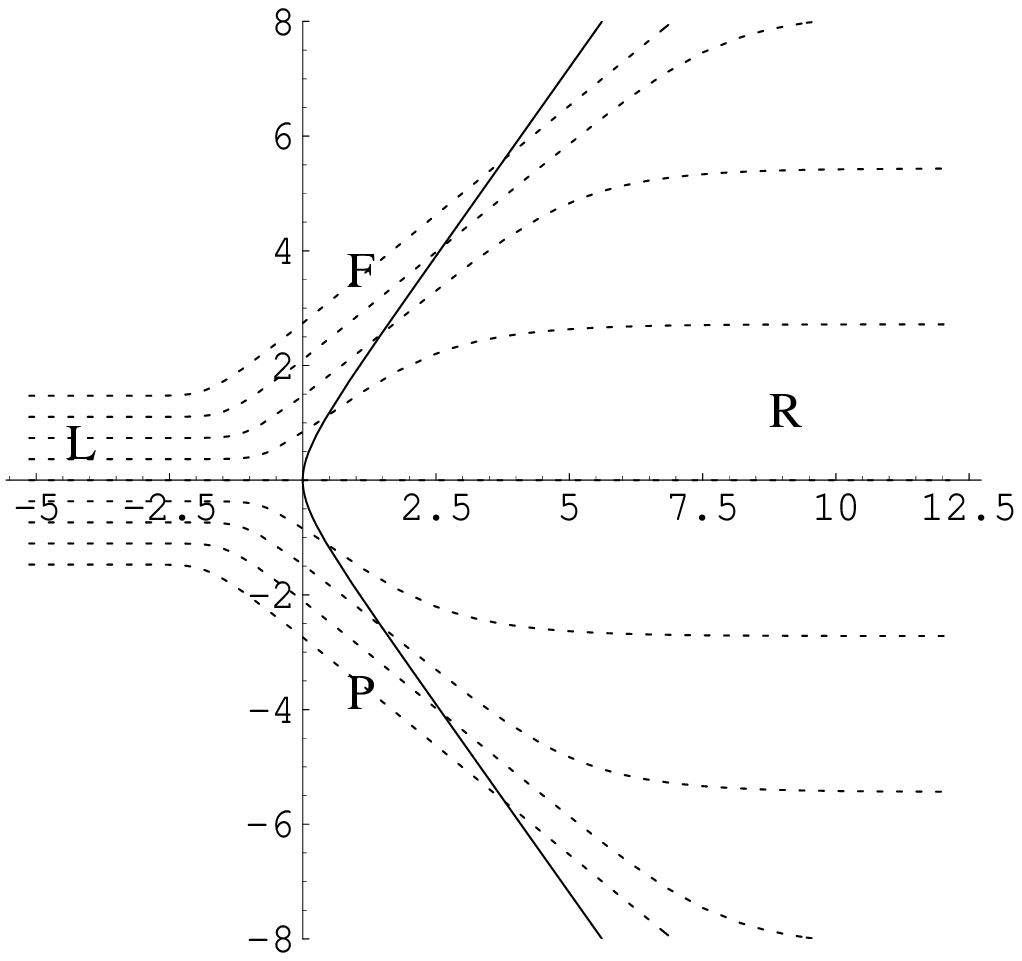, width=6.5cm}

{\footnotesize {\bf Figure \ref{fig2}(A).} Hypersurfaces of simultaneity for the `smooth turnaround' observer in the $(x,t)$ plane (in units of $1/a$) for $\tau_c = 1/a$.}
\end{minipage}\hs{1}\begin{minipage}[b]{6.5cm}
\epsfig{figure=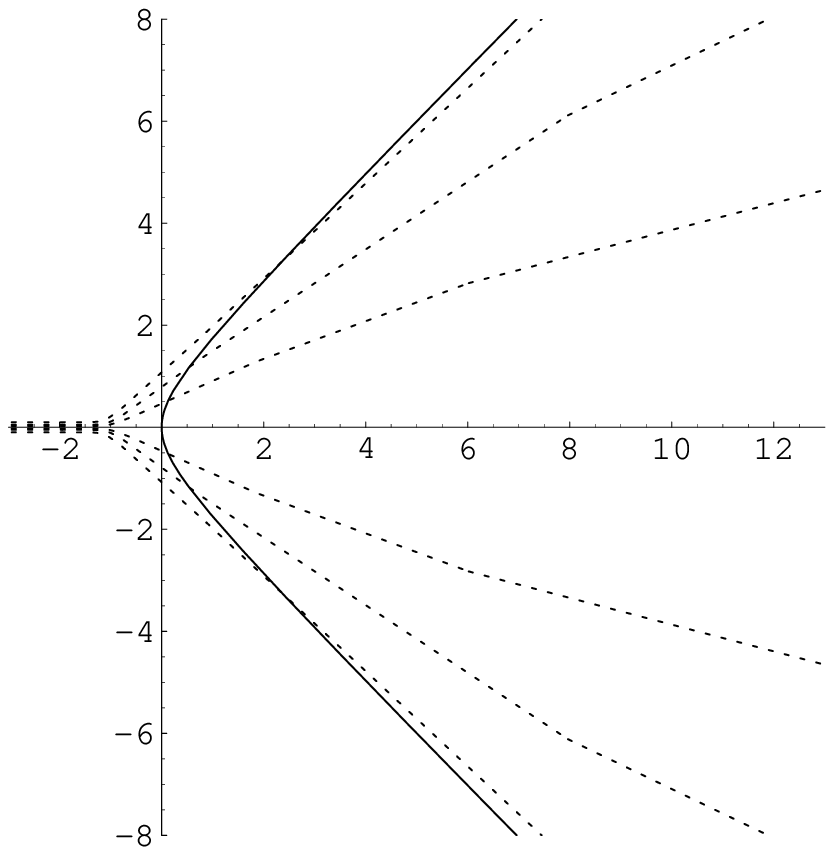, width=6.5cm}

{\footnotesize {\bf Figure \ref{fig2}(B).} Hypersurfaces of simultaneity for the `smooth turnaround' observer, in the $(x,t)$ plane (in units of $1/a$) for $\tau_c = 3/a$.}
\end{minipage}
\figlabel{fig2}
\vs{-.3}
\end{figure}

Figure 2 shows the hypersurfaces of constant $\tau$ for a smooth-turnaround observer (in units of $1/a$) for $\tau_c = 1/a$ (A) and $3/a$ (B). In Figure 2 (A), in which the turn-around time is relatively short, we can distinguish four regions. In region F, where Alice is effectively inertial with velocity $\tanh(a \tau_c)$, her hypersurfaces of `simultaneity' are those appropriate to her status as an inertial observer. The same applies in region P. In regions L, R however, her hypersurfaces of simultaneity are flat, and are contracted/expanded by the Doppler factors $e^{a \tau_c}$ and $e^{-a \tau_c}$ respectively. This prevents the multi-valuedness of the `instantaneous rest frame' construction \cite{Bohm,Dinverno,Yang}. (The point is discussed further in \cite{Dolbyradar}, in the context of the relativistic ``twin-paradox''.) Figure 2 (B) is dominated by the period of acceleration, and bears closer resemblance to the constant acceleration case of Figure 1.

\subsection{Acceleration at Late Times}

Consider an observer (Bob say) with
\begin{equation} \alpha(\tau_{\lambda}) = a \tau_c \log\left(1 + e^{\tau_{\lambda}/\tau_c}\right) \label{latetime1} \end{equation}
His acceleration is: 
\begin{equation} a(\tau_{\lambda}) = \frac{a}{1 + e^{-\tau_{\lambda}/\tau_c}} \label{latetime2} \end{equation}
Bob is stationary at early times, and his acceleration approaches $a$ at late times. Bob's hypersurfaces of simultaneity are shown in Figure 3 (A) (in units of $1/a$) for $\tau_c = 1/a$. There is a future horizon at $x^- = \int_0^{\infty} d \tau \ (1 + e^{\tau/\tau_c})^{-a \tau_c}$, but no past horizon. 

\begin{figure}[htb]
\figstep
\hs{.5}\begin{minipage}[b]{7cm}
\epsfig{figure=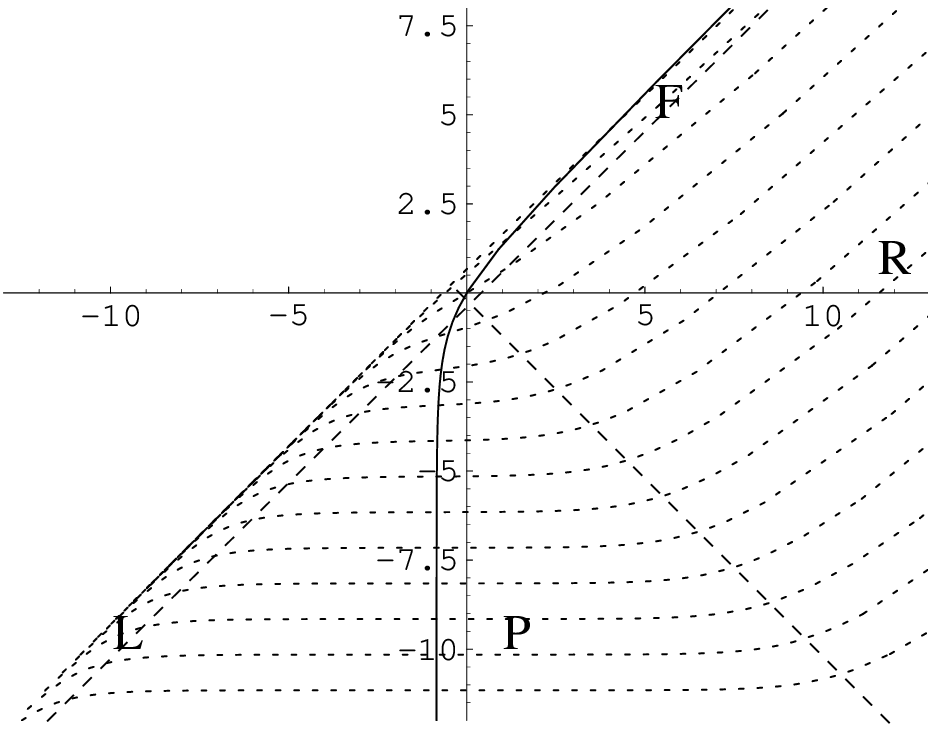, width=7cm}

{\footnotesize {\bf Figure \ref{fig21}(A).} Hypersurfaces of simultaneity for an observer who accelerates at late times, in the $(x,t)$ plane (in units of $1/a$) for $\tau_c = 1/a$.}
\end{minipage}\hs{1}\begin{minipage}[b]{6.5cm}
\epsfig{figure=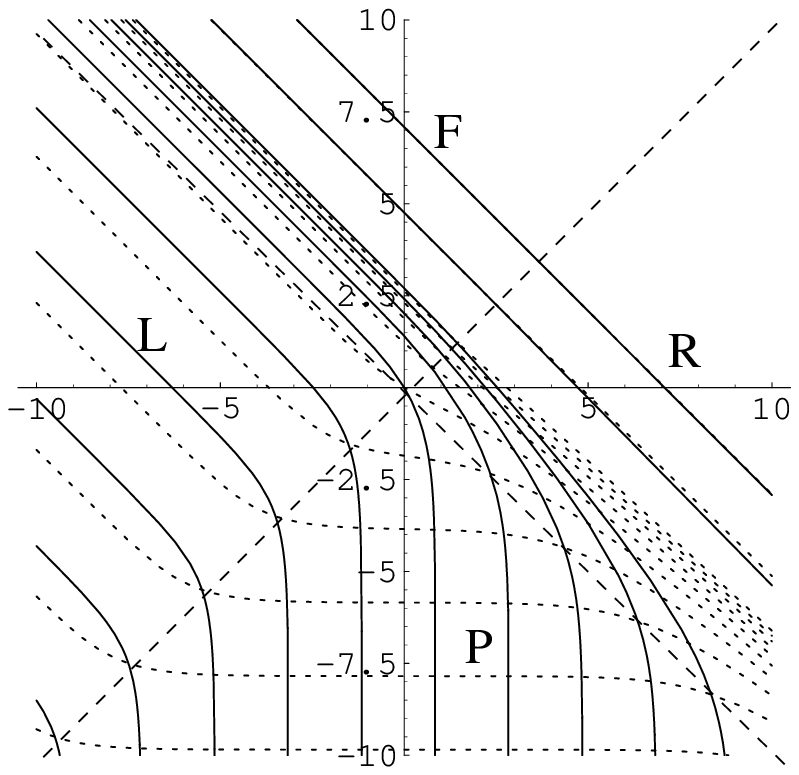, width=6.5cm}

{\footnotesize {\bf Figure \ref{fig21}(B).} Hypersurfaces of constant $t$ (dotted) and constant $x$ (hard) in the $(\rho,\tau)$ plane (in units of $1/a$) of an observer who accelerates at late times, for $\tau_c = 1/a$.}
\end{minipage}
\figlabel{fig21}
\vs{-.3}
\end{figure}

Regions L, F are narrow in Figure 3 (A), and are bounded by the horizon. In the $(\rho,\tau)$ plane, however, describing the space seen by the observer, the horizon is of course not visible (it corresponds to $\tau^- \rightarrow \infty$ for constant $\tau^+$), and the L,F regions are infinite in extent. This can be seen in Figure 3 (B), where surfaces of constant $t$ (dotted) and constant $x$ (hard) are plotted in the $(\rho,\tau)$ plane. The values of $t, x$ shown are $\{-10,-8,-6,-4,-2,0,2,4,6,8,60,600\}$ in units of $1/a$. The line $x=600/a$ (top right) almost coincides with the line $t=600/a$, each being almost null. The same applies to the lines $x=60/a$ and $t=60/a$. All of the $(\rho,\tau)$ plane is covered by $(x,t)$, although the lines of constant $x$ (or of constant $t$) become exponentially tightly packed in regions F and R. The future horizon of Figure 3 (A) implies in Figure 3 (B) that the $x = -4/a$ line (for instance) never crosses the $t = -2/a$ line in the $(\rho,\tau)$ plane. The point $(x,t) = (-4/a,-2/a)$ is outside the observer's causal envelope.

\section{Particle Density of the Dirac Vacuum}

The massless Dirac equation in flat $1 + 1$ dimensional space can be written 
\begin{equation}
\gamma^\mu \frac{\partial \psi}{\partial x^\mu} = 0
\label{diraceqn}\end{equation}
and is frame-invariant provided that the $\gamma^\mu$-matrices  
transform as (contravariant) vectors. By choosing the flat-space Dirac 
matrices $\bar{\gamma}^0 = \left ( \begin{array}{cc} 0 & 1 \\ 1 & 0 \end{array} 
\right )$ and $\bar{\gamma}^1 = \left ( \begin{array}{cc} 0 & -1 \\ 1 & 0 
\end{array} \right )$ (which satisfy $\{\bar{\gamma}^{\mu},\bar{\gamma}^{\nu} \} 
= 2 \eta^{\mu \nu}$) and introducing the basis spinors $\phi_+ = \left 
( 1 \atop 0 \right )$, $\phi_- = \left ( 0 \atop 1 \right )$, we 
can write the general solution as 
\begin{equation}
\psi = \psi_+ (x^-)\phi_+ + \psi_- (x^+) \phi_-
\label{massless} \end{equation}
To define the (inertial) Minkowski vacuum, introduce the plane wave basis:
\begin{equation}
u_{p,F,\pm}  = e^{\mp\imath p x^- }\phi_+, \qquad 
u_{p,B,\pm}  = e^{\mp\imath p x^+ }\phi_- \qquad  \textrm{for } p > 0  
\end{equation}
These modes are normalized to $2\pi \delta (p^\prime - p ) $, and have been decoupled into ``forward-moving" modes (denoted with a subscript ``F"), and ``backward-moving" modes (with subscript ``B").

The Dirac field operator $\hat{\psi}(x)$ can now be written:
\begin{equation}
\hat{\psi}(x) = \int_{0}^{\infty} \frac{d p}{2 \pi} \sum_{\sigma=F,B} \{ u_{p,\sigma,+} a_{p,\sigma} + u_{p,\sigma,-} b^{\dagger}_{p,\sigma} \} \label{fieldop1}\end{equation}
and the Dirac vacuum $|0_{M}\ra$ can be defined by the condition $a_{p,\sigma} | 0_M\ra = 0 = b_{p,\sigma} | 0_M\ra$ for all $p,\sigma$.

	Consider now an arbitrarily moving observer, with coordinates $(\tau,\rho)$. With $\gamma^0 = \frac{\partial \tau}{\partial t} \bar{\gamma}^0 + \frac{\partial \tau}{\partial z} \bar{\gamma}^1$ and $\gamma^1 = \frac{\partial \rho}{\partial t} \bar{\gamma}^0 + \frac{\partial \rho}{\partial z} \bar{\gamma}^1$, equation (\ref{diraceqn}) becomes 
$$g^{0 0} \frac{\partial \psi}{\partial \tau} = - \gamma^0\gamma^1 \frac{\partial \psi}{\partial \rho}$$ 
Since $\gamma^0 \gamma^1 = g^{0 0} \bar{\gamma}^0 \bar{\gamma}^1$, this can be written as

\begin{equation}
i \frac{\partial \psi}{\partial \tau} = - i \bar{\gamma}^0 \bar{\gamma}^1 \frac{\partial \psi}{\partial \rho} \equiv \hat{H}_{nh} \psi
\label{Hnh} \end{equation}

Following \cite{Dolby,Dolby3} we construct particle/antiparticle modes for this observer, by diagonalising the `second quantized' Hamiltonian:

\begin{eqnarray} \hs{-.1} \hat{H}(\tau) & = \int_{\Sigma_{\tau}} \sqrt{-g} T_{\mu \nu}(\hat{\psi}(x)) k^{\mu} d \Sigma^{\nu} \\
 \hs{-1} & = \frac{i}{2} \int \left( \hat{\psi}^{\dagger} M \frac{\partial \hat{\psi}}{\partial \tau} - \frac{\partial \hat{\psi}^{\dagger}}{\partial \tau} M \hat{\psi} \right) d \rho \label{2ndHam}\\
\hs{-1} \mbox{where } \ M & = \sqrt{-g} \bar{\gamma}^0 \gamma^0 = \half \left [ e^{-\alpha(\tau^-)} (\mathbb{I} + \bar{\gamma}^0 \bar{\gamma}^1 ) + e^{\alpha(\tau^+)} (\mathbb{I} - \bar{\gamma}^0 \bar{\gamma}^1 ) \right ] \end{eqnarray}

We now seek the spectrum of the `first quantized' Hamiltonian $\hat{H}_1(\tau) = \half(\hat{H}_{nh} + \hat{H}^{\dagger}_{nh}$), which acts on states defined on $\Sigma_{\tau}$. The inner product expressed on such hypersurfaces can be written in radar coordinates as: 
 \begin{equation}
\langle \psi \vert \phi \rangle_{\Sigma_{\tau_0}} =  \int\limits_{- \infty}^{\infty} d\rho \sqrt{- g}\psi^\dagger \bar{\gamma}^0 \gamma^0 \phi = \int\limits_{- \infty}^{\infty} d\rho \ \psi^\dagger M \phi \label{inner1}\end{equation}
To deduce $\hat{H}^{\dagger}_{nh}$, consider
\begin{eqnarray}
\langle \hat{H}_{nh}\psi\vert \phi \rangle_{\Sigma_{\tau_0}} & = & \frac{1}{2} \int\limits_{-\infty}^{\infty} \! d\rho \left ( \hat{H}_{nh}\psi \right )^{\dagger} M \phi \nonumber \\
& = & \langle \psi\vert -\imath\sigma \frac{\partial}{\partial \rho}\vert \phi \rangle_{\Sigma_{\tau_0}} - \int\limits_{-\infty}^{\infty} d\rho \: \psi^\dagger  M \left (M^{-1} \imath\sigma  \frac{\partial M}{\partial \rho} \right ) \phi 
\end{eqnarray}
where we have discarded surface terms at $\rho \rightarrow \pm \infty$, and have assumed that the functions $U^+ $ and $U^- $ are continuous. This gives our Hermitian Hamiltonian $\hat{H}_1 $ as
\begin{eqnarray}
\hat{H}_1(\tau_0) & = & -i\sigma \frac{\partial}{\partial \rho} \nonumber \\
&   & + \frac{\imath}{4}\left [a(\tau_0 + \rho) (\mathbb{I} - \bar{\gamma}^0 \bar{\gamma}^1) - a(\tau_0 - \rho) (\mathbb{I} + \bar{\gamma}^0 \bar{\gamma}^1)\right ]
\label{ham} \end{eqnarray}
The positive/negative eigenstates of this Hamiltonian will determine the particle/antiparticle content of the Minkowski vacuum measured by our accelerated observer.

By writing these eigenstates as $\psi_{\tau_0} = f_+ \phi_+ + f_- \phi_- $, equation (\ref{ham}) gives
\begin{eqnarray}
\omega f_+ & = & - \imath \frac{\partial f_+ }{\partial \rho }  -  \frac{\imath }{2} a(\tau_0 - \rho) f_+ \nonumber \\
\omega f_- & = & \imath \frac{\partial f_-}{\partial \rho }  +  \frac{\imath}{2} a(\tau_0 + \rho) f_-
\end{eqnarray}

and since $\left ( \frac{\partial}{\partial \rho} \right )_\tau  =  \left ( \frac{\partial}{\partial \tau^+} \right )_\tau = - \left ( \frac{\partial}{\partial \tau^-} \right )_\tau $, these equations can be rewritten as

\begin{eqnarray}
\omega f_+ & = & \imath \frac{df_+ }{d\tau^- }  -  \frac{\imath }{2} a(\tau^-) f_+ \nonumber \\
\omega f_- & = & \imath \frac{df_-}{d\tau^+ }  +  \frac{\imath}{2} a(\tau^+) f_- 
\label{mlesseqns} \end{eqnarray}
with solutions
\begin{eqnarray}
f_+ & = & A_\omega e^{\alpha(\tau^-)/2} e^{-\imath\omega\tau^- } \nonumber \\
f_- & = & B_\omega e^{-\alpha(\tau^+)/2} e^{-\imath\omega\tau^+ }
\end{eqnarray}
where $A_\omega$ and $B_\omega $ are normalisation constants. Comparison with 
equation (\ref{massless}) (and observing that $\tau^+$ is a function of $x^+$, $\tau^-$ a 
function of $x^-$) reveals that in (\ref{mlesseqns}) we have 
found not only eigenstates on $\Sigma_{\tau}$, but have derived solutions 
to the Dirac equation (\ref{diraceqn}) which are eigenstates for all 
$\tau$. This is in general only possible for the massless case, and is a 
consequence of conformal invariance. By choosing $\{A = 1 , B = 0 \}$ and $\{A = 0 , B = 1 \}$ respectively, we can define ``forward-moving'' and ``backward-moving'' modes:

\begin{equation} \hs{-2} \psi_{\omega,F,\pm} = e^{\alpha(\tau^-)/2} e^{\mp\imath \omega \tau^- } \phi_+, \qquad 
\psi_{\omega,B,\pm}  = e^{-\alpha(\tau^+)/2} e^{\mp\imath \omega \tau^+ } \phi_- \qquad  \textrm{for } \omega > 0  
\end{equation} 
These are `normalized' to  $2\pi\delta(\omega - \omega^\prime)$ with 
respect to (\ref{inner1}), but are not yet defined in general over the 
whole spacetime. We wish to write the field operator $\hat{\psi}(x)$ in terms of 
these modes as:
\begin{eqnarray} & \hs{.5} \hat{\psi}(x) \ = \hat{\psi}^{(+)}(x) + \hat{\psi}^{(-)}(x) + \hat{\psi}^{(0)}(x) \label{operator}\\
\hs{-1} \mbox{where } \hs{2} & \hat{\psi}^{(+)}(x) \ \equiv \int \frac{d \omega}{2 \pi} \sum_{\sigma=F,B} \psi_{\omega,\sigma,+} \, \tilde{a}_{\omega,\sigma} \nonumber \\
 & \hat{\psi}^{(-)}(x) \ \equiv \int \frac{d \omega}{2 \pi} \sum_{\sigma=F,B} \psi_{\omega,\sigma,-} \, \tilde{b}^{\dagger}_{\omega,\sigma} \end{eqnarray}
and $\hat{\psi}^{(0)}(x)$ has compact support outside the causal envelope of our observer. To describe this construction, acceleration horizons must be considered. 
Consider the behavior of $x^{\pm}(\tau_{\lambda})$ as $\tau_{\lambda} \rightarrow \infty$. There are three possibilities:
\begin{eqnarray} 
x^+(\tau_\lambda) \rightarrow x^+_{\infty} < \infty & \hs{1} \mbox{and } x^-(\tau_\lambda) \rightarrow \infty \hs{1} & \mbox{(A)} \nonumber \\
x^-(\tau_\lambda) \rightarrow x^-_{\infty} < \infty & \hs{1} \mbox{and } x^+(\tau_\lambda) \rightarrow \infty \hs{1} & \mbox{(B)} \nonumber \\
x^+(\tau_\lambda) \rightarrow \infty & \hs{1} \mbox{and } x^-(\tau_\lambda) \rightarrow \infty \hs{1} & \mbox{(C)} \nonumber 
\end{eqnarray}
($x^{\pm}_{\infty}$ cannot both be finite, since $\frac{d x^+}{d \tau_{\lambda}} \frac{d x^-}{d \tau_{\lambda}} = 1$ for all $\tau_{\lambda}$). Similarly, as $\tau_{\lambda} \rightarrow -\infty$ we have the possibilities
\begin{eqnarray} 
x^+(\tau_\lambda) \rightarrow x^+_{-\infty} > - \infty & \hs{1} \mbox{and } x^-(\tau_\lambda) \rightarrow -\infty \hs{1} & \mbox{(A')} \nonumber \\
x^-(\tau_\lambda) \rightarrow x^-_{-\infty} > - \infty & \hs{1} \mbox{and } x^+(\tau_\lambda) \rightarrow - \infty \hs{1} & \mbox{(B')} \nonumber \\
x^+(\tau_\lambda) \rightarrow - \infty & \hs{1} \mbox{and } x^-(\tau_\lambda) \rightarrow - \infty \hs{1} & \mbox{(C')} \nonumber 
\end{eqnarray}

These combine to give nine possible structures of the causal envelope of the observer. We consider four of these cases here (the other five are similar):

\begin{figure} 
\begin{center}
\epsfig{file=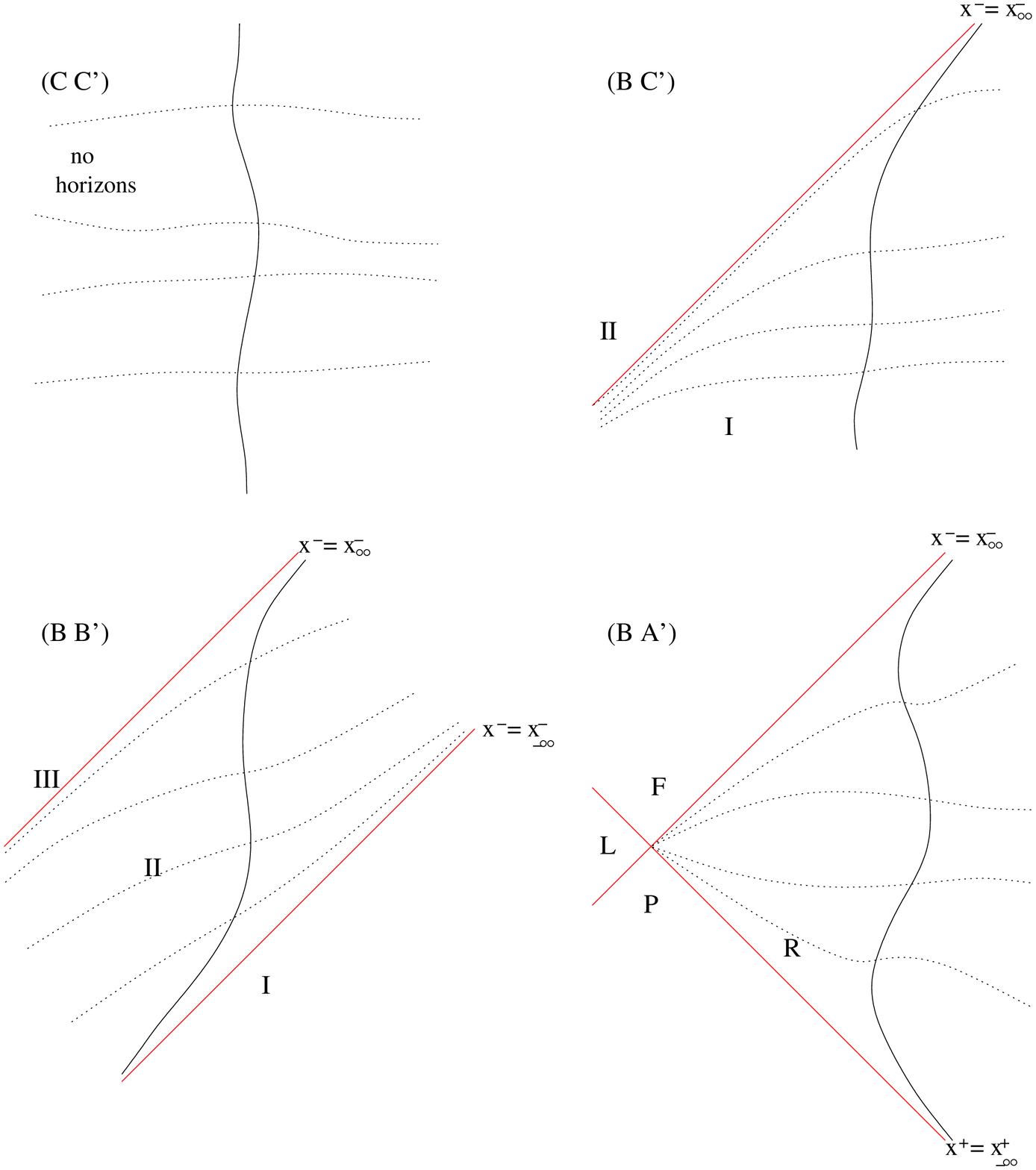,height=4.0in}
\caption{Representation of four of the nine possibilities for the structure of the causal envelope of an observer in 1+1 dimensional flat space. The dark curved lines represent a typical observer from each class, and the dotted lines represent typical `hypersurfaces of simultaneity' for such observers.} \label{fig:cases} 
\end{center}
\end{figure}

\begin{itemize}

\item{(C C')} There are no horizons in this case. It is easy to see that the modes $\psi_{\omega,\sigma,\pm}$ are complete, so that we can achieve (\ref{operator}) with $\hat{\psi}^{(0)}(x) = 0$.

\item{(B C')} In this case there is a future acceleration horizon 
at $x^- = x^-_{\infty}$, which separates the spacetime into two 
regions -- denoted I and II. Region I is the causal envelope of the 
observer, and region II the remainder. The backward-moving modes are 
defined naturally on the whole spacetime, since they are functions 
of $x^+$ only (the range of $x^+(\tau_\lambda)$ covers the whole 
real line). They form a complete set of backward moving modes. Although 
they were normalized on a hypersurface which is not Cauchy, it is 
straightforward to verify that the same norm holds on any Cauchy 
surface. The forward moving modes, on the other hand, are defined 
only in Region I. They can be extended to the whole spacetime by 
specifying $\psi_{\omega,F,\pm}(x) = 0$ throughout region II. This 
ensures that the Dirac equation is satisfied even on the 
horizon (a wave-packet formed from these  
will be continuous on the horizon) and ensures that they are 
normalized to $(2 \pi) \delta(\omega - \omega')$ on any Cauchy 
surface. Although they do not form a complete set, they 
can easily be supplemented by further `forward moving modes', 
$\psi_{i,II}(x)$, which are zero outside region II, and which together 
with the $\{ \psi_{\omega,F,\pm}(x) \}$ comprise a complete set. (The 
$\psi_{i,II}(x)$ might be chosen to be the forward 
moving modes of an observer in Case (C B'), which has $x^-_{\infty}$ 
as a past horizon.) We can then achieve (\ref{operator}) with 
$$\hat{\psi}^{(0)}(x) = \sum_i \psi_{i,II}(x) c_{i}$$ 
where the 
operators $c_i$ satisfy anticommutation relations appropriate to 
the normalization of the $\psi_i$. There is considerable 
freedom in choosing these modes, which do not appear 
in the Hamiltonian $\hat{H}(\tau)$ (they all satisfy 
$\hat{H}_1 \psi_{i,II} = 0$) or in the number density operator 
$\hat{N}(x)$ (defined below) for this observer. These modes 
are `unobservable' by our present observer, since they vanish  
throughout the corresponding causal envelope.

\item{(B B')} In this case there are future and past acceleration 
horizons at $x^- = x^-_{\pm \infty}$, which divide the spacetime 
into regions I, II and III as shown in Figure 4. The backward-moving 
modes again form a complete set, defined on the whole spacetime. The 
forward-moving modes can be extended by specifying 
$\psi_{\omega,F,\pm}(x) = 0$ throughout regions I and III. They may 
be supplemented to form a complete set by including modes $\{ \psi_{i,I}(x) \}$ and 
$\{ \psi_{i',III}(x) \}$, with compact support in regions I and III 
respectively. Equation (\ref{operator}) is then achieved with 
$$\hat{\psi}^{(0)}(x) = \sum_i \psi_{i,I}(x) c_{i,I} + \sum_{i'} \psi_{i',III}(x) c_{i',III}$$

\item{(B A')} This includes the well-studied uniform acceleration case (Section 2.2). It involves a future horizon at $x^- = x^-_{\infty}$ and a past horizon at $x^+ = x^+_{-\infty}$, which intersect at a point, and divide the spacetime into 4 regions, denoted L, R, F and P in Figure 4. All hypersurfaces $\Sigma_{\tau}$ terminate at this point of intersection. In this case the backward-moving modes are defined in Regions R and F, and can be extended to the whole spacetime by specifying $\psi_{\omega,B,\pm}(x) = 0$ throughout regions L and P. They then satisfy $\la \psi_{\omega',B,\lambda'} | \psi_{\omega,B,\lambda}\ra = (2 \pi) \delta_{\lambda \lambda'} \delta(\omega - \omega')$ on any Cauchy surface. They can 
be supplemented by backward-moving modes $\psi_{i,LP}(x)$ having compact support in the union of Regions L and P. The forward-moving modes can be similarly extended, and supplemented by modes $\psi_{i',LF}(x)$ having compact support in the union of Regions L and F. Equation (\ref{operator}) is then achieved with 
$$\hat{\psi}^{(0)}(x) = \sum_i \psi_{i,LP}(x) c_{i,LP} + \sum_{i'} \psi_{i',LF}(x) c_{i',LF}$$

\end{itemize}
Equation (\ref{operator}) has now been confirmed in all cases, and it is routine to substitute $\hat{\psi}(x)$ into equation (\ref{2ndHam}) to obtain the (pre-normal-ordered) Hamiltonian:
\begin{equation} \hat{H}(\tau) = \int \frac{d \omega}{2 \pi} \, \omega \sum_{\sigma=F,B} \{ \tilde{a}^{\dagger}_{\omega,\sigma} \tilde{a}_{\omega,\sigma} - \tilde{b}_{\omega,\sigma} \tilde{b}^{\dagger}_{\omega,\sigma} \} \end{equation}
which has successfully been diagonalised, and does not depend on $\hat{\psi}^{(0)}(x)$. The time-independence of the operators $\tilde{a}_{\omega,\sigma}, \tilde{b}_{\omega,\sigma}$ and of the Hamiltonian $\hat{H}(\tau)$ follows because the solutions $\psi_{\omega,\sigma,\pm}(x)$ are eigenstates of $\hat{H}_1(\tau)$ for all $\tau$ (and with the same eigenvalue $\omega$ for all $\tau$).

	The (second-quantized) total number operator $\hat{N}(\tau)$ \cite{Dolby,Dolby3} can be written in terms of particle/antiparticle density operators $\hat{N}^{\pm}(x)$ as 
\begin{eqnarray}
\hat{N}(\tau) & = & \hat{N}^+ + \hat{N}^- = \int_{-\infty}^{\infty} d \rho \, (\hat{N}^+(x) + \hat{N}^-(x))  \\
\hs{-2.5} \mbox{ where } \hs{1} \hat{N}^+(x) & = & \hat{\psi}^{(+)\dagger}(x) M \hat{\psi}^{(+)}(x) \\
& = & \int\limits_{0}^{\infty} \frac{d \omega}{2 \pi} \int\limits_{0}^{\infty} \frac{d \omega'}{2 \pi} \{ e^{i(\omega'-\omega)\tau^-} \tilde{a}^{\dagger}_{\omega',F} \tilde{a}_{\omega,F} + e^{i(\omega'-\omega)\tau^+} \tilde{a}^{\dagger}_{\omega',B} \tilde{a}_{\omega,B} \} \\
 & = & \hat{N}^+_F(\tau^-) + \hat{N}^+_B(\tau^+) \\
\hs{-2.5} \mbox{ (say), and } \hs{.6} \hat{N}^-(x) & = & - :\hat{\psi}^{(-)\dagger}(x) M \psi^{(-)}(x): \\
& = & \int\limits_{0}^{\infty} \frac{d \omega}{2 \pi} \int\limits_{0}^{\infty} \frac{d \omega'}{2 \pi} \{ e^{i(\omega'-\omega)\tau^-} \tilde{b}^{\dagger}_{\omega',F} \tilde{b}_{\omega,F} + e^{i(\omega'-\omega)\tau^+} \tilde{b}^{\dagger}_{\omega',B} \tilde{b}_{\omega,B} \} \\
 & = & \hat{N}^-_F(\tau^-) + \hat{N}^-_B(\tau^+) \end{eqnarray}
where the normal ordering is with respect to the observer's particle interpretation (i.e. the $\tilde{b}_{\omega,F}$). By equating expressions (\ref{fieldop1}) and (\ref{operator}) for $\hat{\psi}(x)$, we find that

\begin{eqnarray} \tilde{a}_{\omega,\sigma} & = \int_{0}^{\infty} \frac{d p}{2 \pi} \{ \la \psi_{\omega,\sigma,+}|u_{p,\sigma,+}\ra \, a_{p,\sigma} + \la \psi_{\omega,\sigma,+}|u_{p,\sigma,-}\ra \, b^{\dagger}_{p,\sigma} \} \\
 \tilde{b}_{\omega,\sigma} & = \int_{0}^{\infty} \frac{d p}{2 \pi} \{ \la \psi_{\omega,\sigma,-}|u_{p,\sigma,+}\ra^* \, a^{\dagger}_{p,\sigma} + \la \psi_{\omega,\sigma,-}|u_{p,\sigma,-}\ra^* \, b_{p,\sigma} \} \end{eqnarray}
(A similar expression exists for the $\tilde{c}_i$.) Substitution into $\hat{N}^{\pm}$ allows us to evaluate the particle content of the Dirac Vacuum $|0_M\ra$ as measured by this observer. For instance, the total number of particles/antiparticles measured is:

\begin{eqnarray}
N^+ & = \la 0_M|\hat{N}^+|0_M\ra = & \int\limits_0^{\infty} \frac{d\omega}{2\pi} \sum_{\sigma=F,B} (\beta\beta^\dagger )_{\omega\omega,\sigma}  \nonumber \\
N^- & = \la 0_M|\hat{N}^-|0_M\ra = & \int\limits_0^{\infty} \frac{d\omega}{2\pi} \sum_{\sigma=F,B} (\gamma\gamma^\dagger )_{\omega\omega,\sigma} 
\label{totalps} \end{eqnarray}
where 
\begin{eqnarray}
\beta_{\omega p, F} & \equiv & \la \psi_{\omega,F,+}|u_{p,F,-}\ra =\int\limits_{-\infty}^{\infty} d\rho \: e^{-\alpha(\rho)/2} e^{\imath\omega\rho} e^{\imath p x^-_{\lambda}(\rho) }  \nonumber \\
\beta_{\omega p, B} & \equiv & \la \psi_{\omega,B,+}|u_{p,B,-}\ra =\int\limits_{-\infty}^{\infty} d\rho \: e^{\alpha(\rho)/2} e^{\imath\omega\rho} e^{\imath p x^+_{\lambda}(\rho) } 
\label{betas} \\
\hs{-1.9} \mbox{ and } \hs{1} \gamma_{\omega p, \sigma} & \equiv & \la \psi_{\omega,\sigma,-}|u_{p,\sigma,+}\ra  =  \beta_{\omega p, \sigma}^{*} \label{partvanti}
\end{eqnarray}
Here $\beta_{\omega p, \sigma}$ and $\gamma_{\omega p, \sigma}$ are independent of $\tau$, since the states $\psi_{\omega,\sigma,\pm}$ are all solutions of the Dirac equation. The variable $\rho$ serves simply as an integration dummy. Though its physical origins come from an integral over space we could equally well use the label $\tau_{\lambda}$, viewing the integral as an integral along the observer's worldline.

From equation (\ref{partvanti}) we see that $(\gamma\gamma^\dagger )_{\omega\omega,\sigma} = (\beta\beta^\dagger )_{\omega\omega,\sigma}$ for both $\sigma$, so that from (\ref{totalps}) the total number and frequency distribution of `forward'-moving particles are equal to those of `forward'-moving antiparticles, and similarly for `backward'-moving particles. In general, the forward-moving particle content need not resemble the backward-moving particle content.

The formula for the distribution of particles also takes a simple form:
\begin{eqnarray}
n^+_F (\tau, \rho) & \equiv \la 0_M | \hat{N}^+_F(x) |0_M\ra = &  \int\limits_{0}^{\infty}\frac{d\omega}{2\pi}\int\limits_{0}^{\infty}\frac{d\omega^\prime}{2\pi}(\beta\beta^\dagger )_{\omega\omega^\prime\!\!,\, F} \ e^{\imath (\omega^\prime - \omega )\tau^-}  \nonumber \\
n^-_F (\tau, \rho) & \equiv \la 0_M | \hat{N}^-_F(x) |0_M\ra = &  \int\limits_{0}^{\infty}\frac{d\omega}{2\pi}\int\limits_{0}^{\infty}\frac{d\omega^\prime}{2\pi}(\gamma\gamma^\dagger )_{\omega\omega^\prime\!\!,\, F} \ e^{- \imath (\omega^\prime - \omega )\tau^-}  \label{density} \\
& = n^+_F (\tau, \rho)
\end{eqnarray}
where equation (\ref{partvanti}) has been used to reach the last line. The spatial distributions of particles and antiparticles are the same, as expected from local conservation of charge. Further, the integral over $p>0$ which is implicit in the definition of $(\beta \beta^{\dagger})_{\omega\omega',\sigma}$ can be evaluated explicitly to yield:
\begin{eqnarray} (\beta \beta^{\dagger})_{\omega \omega',F} & = \frac{2}{\pi} \int_{-\infty}^{\infty} {\rm d} \tau_a e^{-2 i \omega_d \tau_a} \int_{0}^{\infty} {\rm d} \tau_d \sin(2 \omega_a \tau_d) g_{+}(\tau_a,\tau_d) \\
\hs{-1} \mbox{ where } \hs{1} \omega_a & = \half(\omega + \omega') \hs{1} \omega_d = \half(\omega' - \omega) \\
\hs{-1} \mbox{ and } \ \ g_{\pm}(\tau_a,\tau_d) & = \frac{1}{2 \tau_d} - 
\frac{\exp[\frac{\mp 1}{2} (\alpha(\tau_a + \tau_d) + \alpha(\tau_a - \tau_d))]}{\int_{- \tau_d}^{\tau_d} {\rm d} \tau \, \exp[\mp\alpha(\tau_a + \tau)] } \label{gfunction} \end{eqnarray}

In equation (\ref{density}) this yields:
\begin{equation} n^{\pm}_{F}(\tau^-) = \frac{1}{4 \pi^2} \int_{-\infty}^{\infty} \frac{{\rm d} \tau}{\tau} g_+(\tau^- + \tau,\tau) \hs{.5} = \int_{0}^{\infty} \frac{{\rm d} \omega_a}{2 \pi} n_{F,\omega}(\tau^-) \label{spatialdist} \end{equation}
where 
\begin{equation} n_{F,\omega}(\tau^-) = \frac{2}{\pi^2} \int_{-\infty}^{\infty} {\rm d} \tau_a \ 
\frac{\sin[2 \omega (\tau^- - \tau_a)]}{\tau^- - \tau_a} 
\int_{0}^{\infty} {\rm d} \tau_d \ \sin(2 \omega \tau_d) g_{+}(\tau_a,\tau_d) 
\label{freqdist}\end{equation}

This density is a function only of $\tau^- = \tau-\rho$, as expected for 
forward-moving massless particles. It is defined such that $n^{\pm}_{F}(\tau^-) {\rm d} \rho$ 
gives the number of particles within ${\rm d} \rho$ of the point $(\tau,\rho)$.
Integration of $n_{F,\omega}(\tau^-)$ over $\rho$ at any time $\tau$ gives the total number of forward-moving particles of frequency $\omega$:
\begin{equation} (\beta \beta^{\dagger})_{\omega \omega,F} = \int_{-\infty}^{\infty}  d \rho \, n_{F,\omega}(\tau^-) \hs{.2} = \int_{-\infty}^{\infty}  d \tau^- \, n_{F,\omega}(\tau^-)\end{equation}

In combination with equation (\ref{spatialdist}), this result suggests that 
$n_{F,\omega}(\tau^-)$ should be interpreted 
as the frequency distribution of forward moving 
particles/antiparticles at the point $\tau^-$. Though this interpretation is 
reasonable, $n_{F,\omega}(\tau^-)$ must be averaged over a suitable range of $\omega$ and $\rho$  before its predictions become more significant than the fluctuations \cite{Dolby3}. This limitation is manifest for instance, in the fact that $n_{F,\omega}(\tau^-)$ is not positive definite, whereas $n^{\pm}_{F}(\tau^-)$ and $(\beta \beta^{\dagger})_{\omega \omega,F}$ are positive definite.

Equation (\ref{freqdist}), together with (\ref{gfunction}), specifies the 
distribution $n_{F,\omega}(\tau^-)$ anywhere in the 
spacetime directly in terms of the observer's rapidity. Similarly, the spatial 
distribution of backward-moving particles/antiparticles is given by:

\begin{eqnarray}
n^\pm_B (\tau, \rho)  & =   \int\limits_{0}^{\infty}\frac{d\omega}{2\pi}\int\limits_{0}^{\infty}\frac{d\omega^\prime}{2\pi}(\beta\beta^\dagger )_{\omega\omega^\prime\!\!,\, B} \ e^{\imath (\omega^\prime - \omega )\tau^+} \\
& = \frac{1}{4 \pi^2} \int_{-\infty}^{\infty} \frac{d \tau}{\tau} g_-(\tau^+ + \tau,\tau) \hs{.5} = \int_0^{\infty} \frac{{\rm d} \omega}{2 \pi} n_{B,\omega}(\tau^+) \label{density2} \end{eqnarray}
where 
\begin{equation} n_{B,\omega}(\tau^+) = \frac{2}{\pi^2} \int_{-\infty}^{\infty} {\rm d} \tau_a \ \frac{\sin[2 \omega (\tau^+ - \tau_a)]}{\tau^+ - \tau_a} 
\int_{0}^{\infty} {\rm d} \tau_d \ \sin(2 \omega \tau_d) g_{-}(\tau_a,\tau_d) 
\label{freqdist2} \end{equation}
	Clearly the backward moving particle content (which exactly 
matches the backward moving antiparticle content) is a function only 
of $\tau^+ = \tau + \rho$.

	Although the distributions of `forward'- and `backward'-moving particles are not generally the same, we might expect them to be related when the observer's trajectory is time-symmetric. To study this possibility, consider an observer whose trajectory satisfies the time-reversal symmetry
\begin{eqnarray}
x_\lambda (-\tau_\lambda ) & = & x_\lambda (\tau_\lambda ) \nonumber \\
t_\lambda (-\tau_\lambda ) & = & - t_\lambda (\tau_\lambda ) 
\label{timerev} \end{eqnarray}
It follows that $\alpha(-\tau_\lambda) = - \alpha(\tau_{\lambda})$, which gives 
$n_{F,\omega}(\tau') = n_{B,\omega}(-\tau')$ for all $\tau'$. On the 
hypersurface $\tau=0$, for instance, where $\tau^{\pm} = \pm \rho$, this 
implies that the distribution of forward moving particles exactly matches 
that of backward moving particles. However, on the observer's worldline, the 
distribution of forward moving particles is the time-reverse of 
the distribution for backward moving particles.

\section{Examples}

\subsection{Inertial Observer}
\label{init}

For an inertial observer, $\alpha$ is constant, which gives $g_{\pm}(\tau_a,\tau_d) = 0$. The particle content 
is therefore everywhere zero, as expected.

\subsection{Constant Acceleration}

For a uniformly accelerating observer we have
$$ g_+(\tau_a,\tau_d) = g_-(\tau_a,\tau_d) = \frac{1}{2 \tau_d} - \frac{a}{2 \sinh(a \tau_d)} $$
which is independent of $\tau_a$. Hence the forward and backward moving 
particledistributions are both uniform w.r.t. $\rho$ for all $\tau$, and the 
frequency distribution is everywhere given by:

\begin{eqnarray} n_{F,\omega} & = n_{B,\omega} = \frac{1}{\pi} \int_0^{\infty} d \tau_d \ \sin(2 \omega \tau_d) \left( \frac{1}{\tau_d} - \frac{a}{\sinh(a \tau_d)}\right) \\
& = \frac{1}{1 + e^{\frac{2 \pi \omega}{a}}} \label{thermal} \end{eqnarray}
which is a thermal spectrum at temperature $T = \frac{a}{2 \pi k_B}$, as expected \cite{Takagi,Soffel,QEDSF}.

\begin{figure}
\center{\epsfig{figure=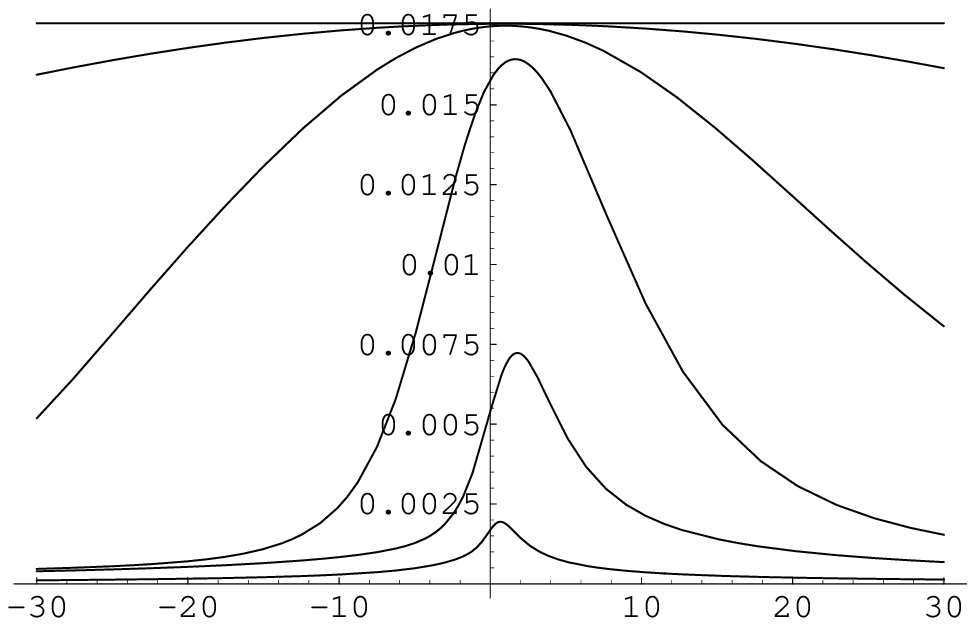, width=8cm}}
\caption{{\footnotesize $n_{F,B}(\rho)/a$ for the smooth turnaround observer, as a function of $a \rho$ for $\tau=0$ and $a \tau_c = 1$ (bottom curve), $3, 10, 30, 100$ and $\infty$ (top curve).}}
\end{figure}

\subsection{A `Smooth Turnaround' Observer.}

Consider again the observer with acceleration 
$$a(\tau_{\lambda}) =  a \cosh^{-2}\left(\frac{\tau_{\lambda}}{\tau_c}\right)$$
By substituting the corresponding rapidity $\alpha(\tau_{\lambda}) = a \tau_c \tanh(\tau/\tau_c)$ into equation (\ref{gfunction}), we immediately 
obtain the spatial distribution of forward or backward moving particles. At 
time $\tau=0$ these distributions are equal, as a result of the time-reversal invariance of the observer's trajectory. They are shown in Figure 5 as 
a function of $a \rho$ for $a \tau_c = 1$ (bottom curve), $3, 10, 30, 100$ and $\infty$ 
(top line). As $\tau_c$ increases the particle density increases, and approaches 
the spatial uniformity found for the `constant-acceleration' ($\tau_c \rightarrow \infty$) limit.

\begin{figure}[htb]
\figstep
\hs{.5}\begin{minipage}[b]{7cm}
\epsfig{figure=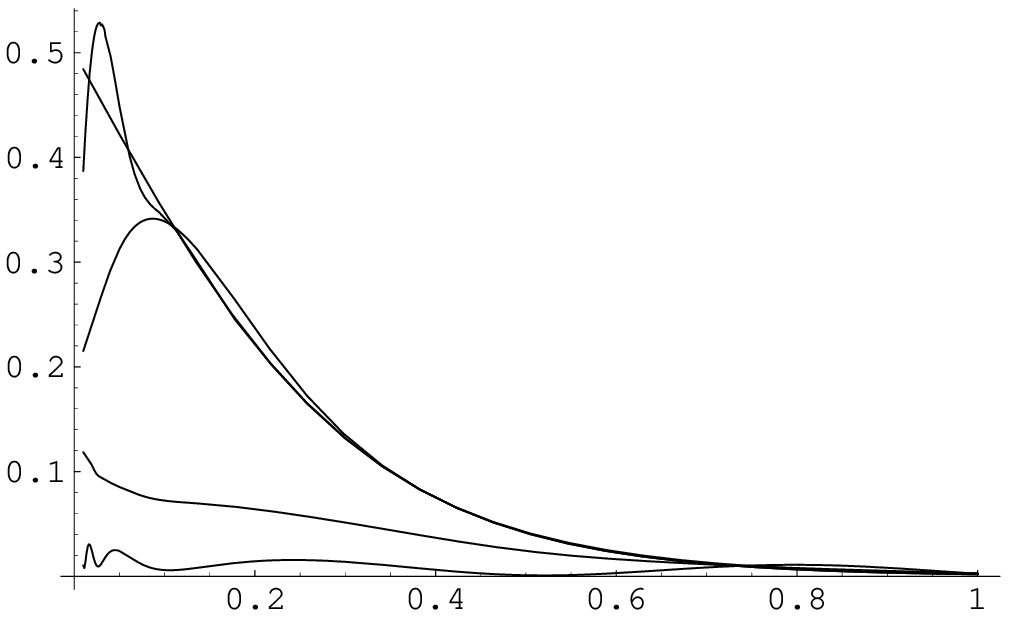, width=7cm}

{\footnotesize {\bf Figure \ref{fig3}(A).} $n_{F,\omega}(\rho)$ for the smooth turnaround observer, as a function of $\omega/a$ for $\rho = 0 = \tau$, and $a \tau_c = 3, 10$ and $30$.}
\end{minipage}\hs{1}\begin{minipage}[b]{7cm}
\epsfig{figure=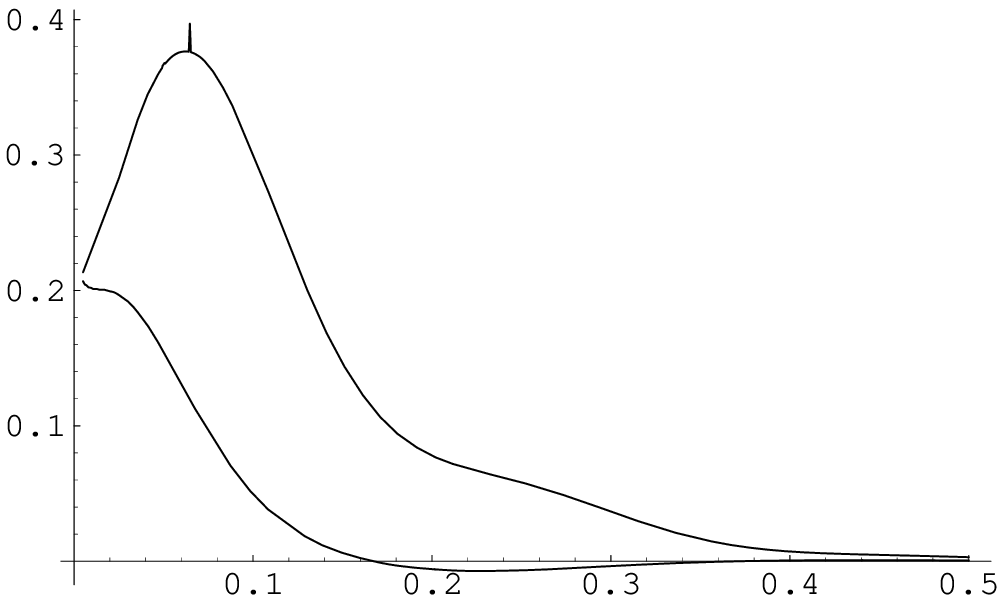, width=7cm}

{\footnotesize {\bf Figure \ref{fig3}(B).} $n_{F,\omega}(\rho)$ for the smooth turnaround observer, as a function of $\omega/a$ for $\tau$, $a \tau_c = 10$ and $a \rho = \pm 10$.}
\end{minipage}
\figlabel{fig3}
\vs{-.3}
\end{figure}

Figure 6 shows the frequency distribution $n_{F,\omega}(\rho)$ of 
forward-moving particles as a function of $\omega/a$ for $\tau=0$. (This also 
equals the backward-moving distribution $n_{B,\omega}(\rho)$ there.) In Figure 
6 (A) we have $\rho=0$ and $a \tau_c = 3, 10, 30$ and $\infty$. We see clearly 
that the distribution approaches the thermal form as $\tau_c$ increases, with only 
the low frequencies differing significantly from the constant acceleration case. In Figure 
6 (B), $a \tau_c = 10$ and $a \rho = + 10$ (top curve) and $-10$ (bottom curve).  
Since $n_{F,B}$ depend only on $\tau^{+}$ and $\tau^{-}$ respectively, Figure 6 (B) also 
represents the distribution of 
forward/backward moving particles on the observer's worldline, at 
$\tau_{\lambda} = 10/a$. At this time the observer sees more backward moving particles than forward moving particles, while the 
forward/backward moving distributions are the opposite at $\tau=-10/a$.

\subsection{Acceleration at Late Times.}

\begin{figure}
\center{\epsfig{figure=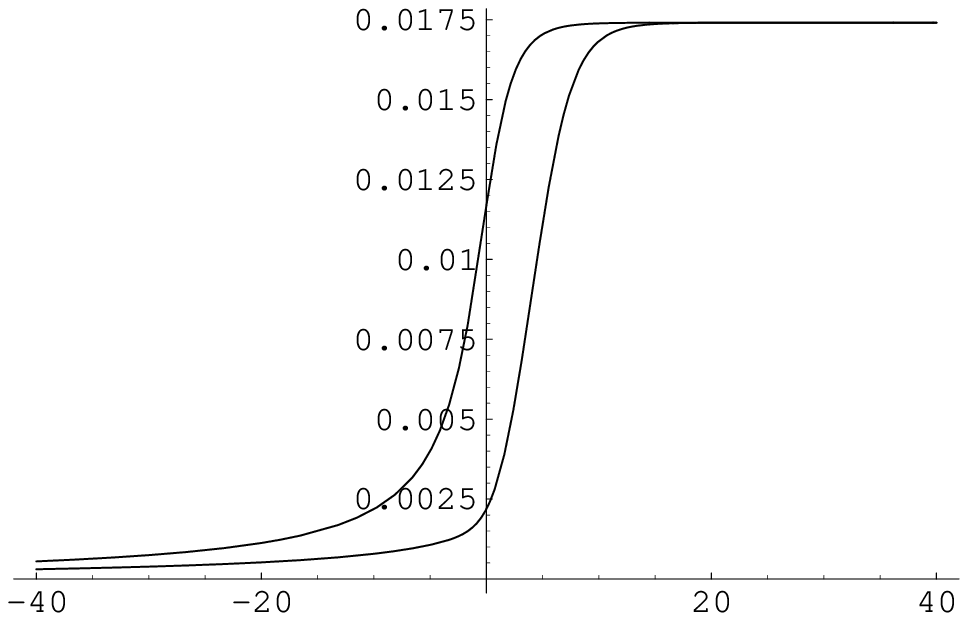, width=8cm}}
\caption{{\footnotesize $n_{F}(\tau)/a$ (top) and $n_{B}(\tau)/a$ (bottom) for 
the `late-time acceleration' observer as a function of $a \tau$ for $\rho=0$ and 
$a \tau_c = 1$.}}
\end{figure}

Consider again the observer of Section 2.4, whose worldline is shown in Figure 3, in 
the $(x,t)$ and $(\rho,\tau)$ planes. Figure 7 shows the spatial distribution of 
forward and backward moving particles on this observer's trajectory. On the trajectory 
the observer consistently sees more forward than backward moving particles. Since 
$n_{F,B}$ depend only on $\tau^{+}$ and $\tau^{-}$ respectively we can again extrapolate from 
Figure 7 to the whole $(\rho,\tau)$ plane. Well into region F (see Figure 3 (B)) 
the forward/backward distributions are approximately uniform, in agreement 
with the predictions of the constant acceleration case. In region L there are  
many forward moving particles but few backward movers. In region R there are many 
backward moving particles but few forward movers, while in region P, where the observer 
is almost inertial, few particles of any variety will be observed. The distributions 
$n_{F,B}(\tau)$ of Figure 7 can be decomposed into their frequency components by plotting 
$n_{B,\omega}(\tau)$ and $n_{F,\omega}(\tau)$ for various values of $\omega$. This 
is done in Figure 8 for $\omega/a = 0.05$ (A) and $0.5$ (B).

\begin{figure}[htb]
\figstep
\hs{.5}\begin{minipage}[b]{7cm}
\epsfig{figure=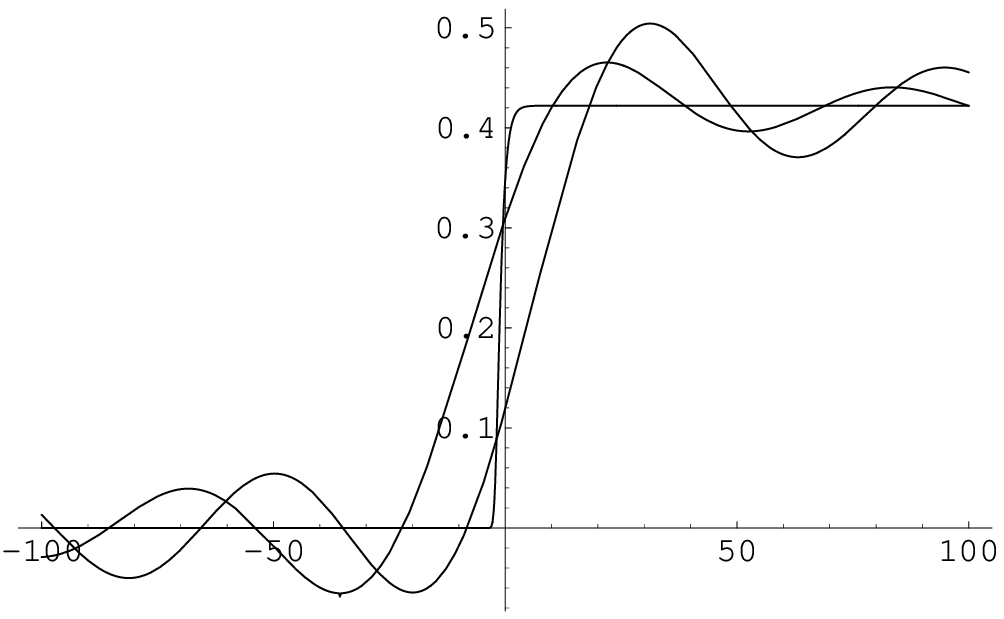, width=7cm}

{\footnotesize {\bf Figure \ref{fig8}(A).} $n_{F,\omega}(\tau)$ and $n_{B,\omega}(\tau)$ for the `late-time acceleration' observer as a function of $a \tau$ for $\rho=0$, $\omega/a = 0.05$ and $a \tau_c = 1$.}
\end{minipage}\hs{1}\begin{minipage}[b]{7cm}
\epsfig{figure=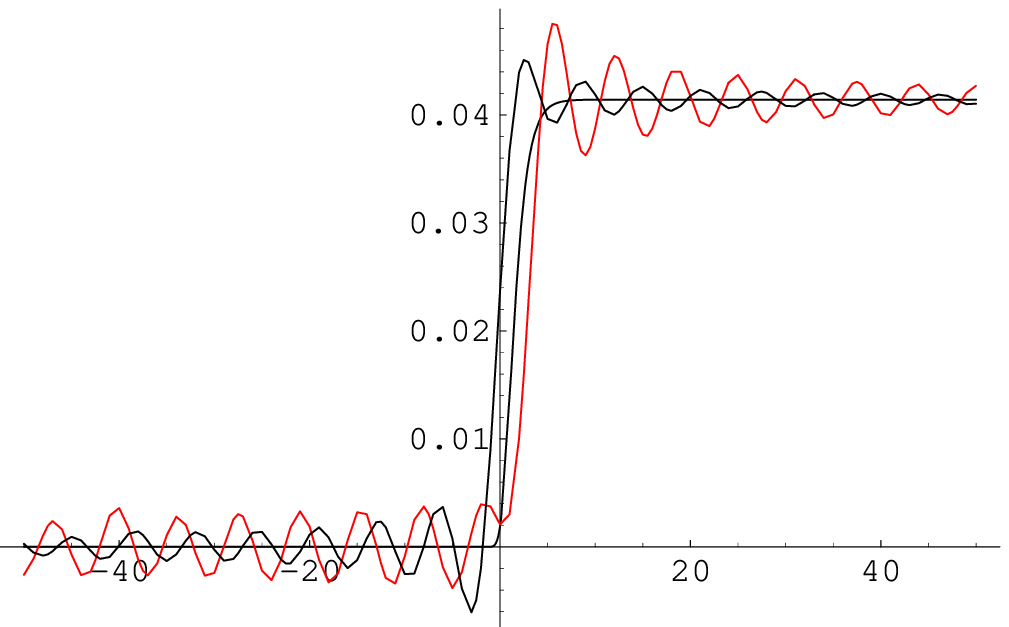, width=7cm}

{\footnotesize {\bf Figure \ref{fig8}(B).} $n_{F,\omega}(\tau)$ and $n_{B,\omega}(\tau)$ for the `late-time acceleration' observer as a function of $a \tau$ for $\rho=0$, $\omega/a = 0.5$ and $a \tau_c = 1$.}
\end{minipage}
\figlabel{fig8}
\vs{-.3}
\end{figure}

These plots clearly show that $n_{F,\omega}(\tau)$ and $n_{B,\omega}(\tau)$ are not positive definite, 
despite the positive definiteness of $n_{F,B}(\tau)$. Also included in Figure 8 are  
the predictions obtained from the naive `instantaneous thermal spectrum' 
$2 \pi/(1 + e^{2 \pi \omega/a(\tau)})$ that would result from substituting 
$a(\tau)$ directly into equation (\ref{thermal}). The high frequency modes, which 
probe only short distance scales, closely trace these naive predictions, whereas the 
low frequency modes take more time to adjust from the `inertial' to the `uniform 
acceleration' predictions.

\begin{figure}[htb]
\figstep
\hs{.5}\begin{minipage}[b]{7cm}
\epsfig{figure=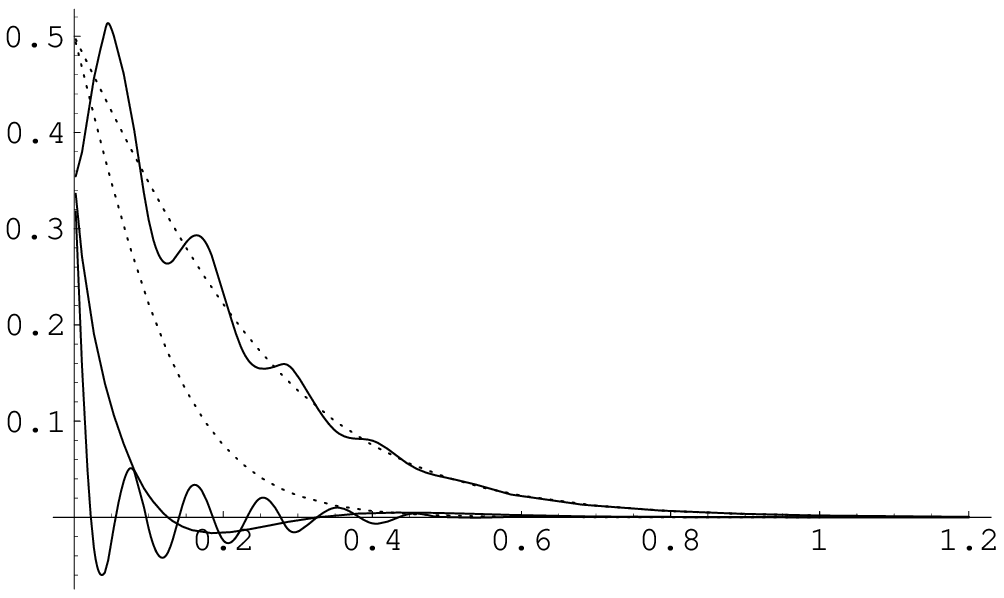, width=7cm}

{\footnotesize {\bf Figure \ref{fig9}(A).} $n_{B,\omega}(\tau)$ for the `late-time acceleration' observer, as a function of $\omega/a$, for $\rho=0$, $a \tau_c = 1$ and $a \tau = -30, 0$ and $30$. Also included are the `instantaneous thermal spectrum' predictions $(1 + e^{2 \pi \omega/a(\tau)})^{-1}$ (dotted).}
\end{minipage}\hs{1}\begin{minipage}[b]{7cm}
\epsfig{figure=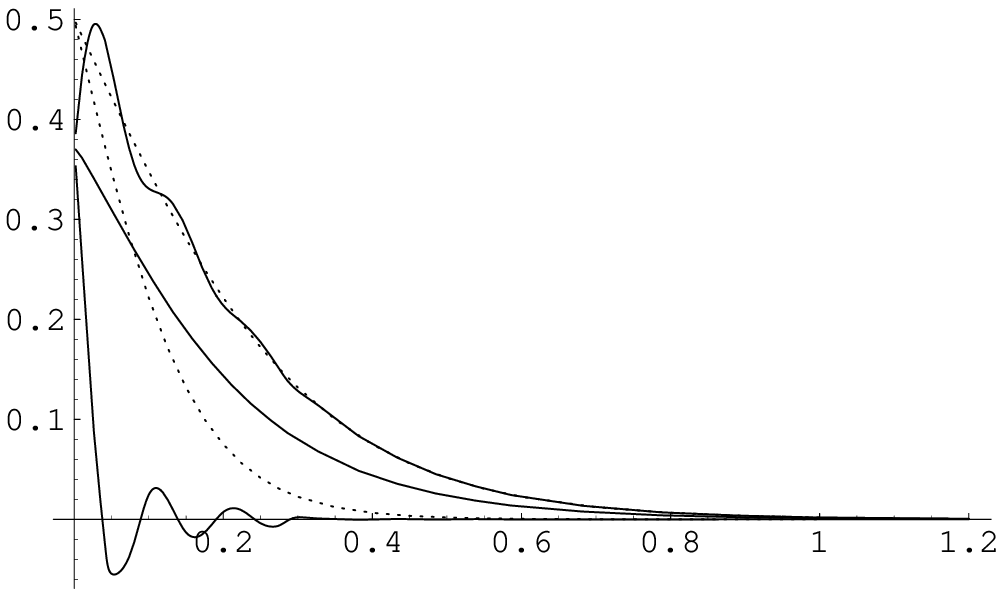, width=7cm}

{\footnotesize {\bf Figure \ref{fig9}(B).} $n_{F,\omega}(\tau)$ for the `late-time acceleration' observer, as a function of $\omega/a$, for $\rho=0$, $a \tau_c = 1$ and $a \tau = -30, 0$ and $30$. Also included are the `instantaneous thermal spectrum' predictions $(1 + e^{2 \pi \omega/a(\tau)})^{-1}$ (dotted).}
\end{minipage}
\figlabel{fig9}
\vs{-.3}
\end{figure}

Figure 9 shows $n_{F,\omega}(\tau)$ and $n_{B,\omega}(\tau)$ as a function of $\omega/a$, for $\tau = -30/a, 0$ and $30/a$. At late times the spectrum is close to thermal. For $\tau = 0$ neither the forward (Figure 9 (B)) nor backward (Figure 9 (A)) moving distributions resemble a thermal spectrum, even at large frequencies, and $n_{B,\omega}(\tau)$ is not positive definite. At much earlier times both distributions are very small, and neither is positive definite. The limit as $\omega \rightarrow 0$ of these distributions is independent of $\tau$, reflecting the fact that the $\omega=0$ mode is a global feature. However, this limit is different for forward and backward moving particles.

\section{Discussion}

The approach developed in \cite{Dolby2,Dolby,Dolby3} has been used to evaluate the 
particle content of the massless Dirac vacuum for an arbitrarily moving 
observer in 1+1 dimensions. This method uses Bondi's `radar time', which 
provides a foliation of flat space that depends only on the observer's 
trajectory. It agrees with proper time on the trajectory, and is single 
valued in the causal envelope of the observer's worldline. 

The particle content can be described in terms of `forward moving' particles 
(with particle density $n^+_F(\tau,\rho) $ dependent only on $\tau - \rho$), 
`backward moving' particles (with particle density
$n^+_B(\tau,\rho)$ dependent only on $\tau + \rho$), and their corresponding 
antiparticles (described by $n^-_F$ and $n^-_B $). The calculation of 
$n^{\pm}_{B,F}(\tau,\rho)$ is largely independent of the possible presence 
of horizons, although when horizons exist an observer cannot assign 
particle/antiparticle densities to regions of spacetime invisible 
to this observer. This is entirely reasonable, though it differs from 
other approaches \cite{Yang}. The particle content is clearly different for 
different observers, but for any given observer the total number of 
forward/backward moving particles is independent of $\tau$. This result 
is specific to massless particles in 1+1 dimensions, and is a direct 
consequence of conformal invariance. The spatial distribution of forward 
moving particles always matches that of forward moving antiparticles, and 
likewise for backward moving particles. In general, there is no connection 
between the observed numbers of forward moving particles and backward moving 
particles. However, when the observer's trajectory is time-reversal invariant, 
the distribution of forward and backward moving particles must match at 
$\tau=0 $. Their distributions at other times is then specified by the fact 
that $n^{\pm}_F(\tau,\rho) $ depends only on $\tau - \rho $ and that 
$n^{\pm}_B(\tau,\rho) $ depends only on $\tau + \rho$. 

These general results 
have been illustrated by four examples. We have confirmed that inertial 
observers do not observe particles in the inertial vacuum, and that uniformly 
accelerating observers measure the well-known Fermi-Dirac spectrum 
\cite{Soffel,Hughes,CaDe,IyKu,Hacyan,Takagi,Horibe,Dolby3,QEDSF}. This thermal 
spectrum is spatially 
uniform in `radar distance' $\rho $ and is the same for forward and backward 
moving particles, as expected. An observer with acceleration 
$a(\tau_{\lambda}) = a \cosh^{-2}(\tau/\tau_c)$ has also been studied. For large 
$\tau_c$, this observer increasingly resembles the uniformly accelerating 
observer, although no horizons are present. For $|\rho|,|\tau| << \tau_c$ 
the particle content measured by this observer closely resembles the 
uniformly accelerating case, particularly for high frequency modes, which 
probe only short distance scales. This sheds light on a common misconception 
-- that the Unruh effect is ``due to the presence of a horizon''. As a result of the 
time-reversal invariance of this example, we find 
$n^{\pm}_F(0,\rho) =  n^{\pm}_B(0,\rho)$ for all $\rho$ as generally predicted, 
although $n^{\pm}_F(\tau,\rho) \neq  n^{\pm}_B(\tau,\rho)$ for $\tau \neq 0$.

	The further example treated here is an observer with $a(\tau_{\lambda}) = \frac{a}{1 + e^{\left(\tau/\tau_c\right)}}$. This observer accelerates uniformly at late times, but is inertial at early times. The observed particle content closely resembles the uniform acceleration result in the causal future of the acceleration, but approaches zero at early times (and away from the horizon). The distributions of forward and backward moving particles now differ significantly, with $n^{\pm}_F(\tau,0) >  n^{\pm}_B(\tau,0)$ for all $\tau$. This intriguing result suggests an observer-dependent violation of discrete symmetries, which calls for further investigation. 

	Since massless fermions in 1+1 dimensions are conformally invariant, and consequently lack the scale dependence of the massive case (provided for instance by the Compton scale $1/m $). The results obtained here will not generalize immediately to massive particles, or to 3+1 dimensions. Further work is needed to investigate those cases in more detail, and to explore the possible observer dependence of discrete symmetries.


\section{References}

\begin{thebibliography}{00}
\bibitem{Un} 
W. G. Unruh, \emph{Phys. Rev. D} \textbf{14} (1976), 870 - 892.
\bibitem{Davies}
P.C.W. Davies, {\it J. Phys. A.} {\bf 8(4)} (1975), 609 - 616.
\bibitem{Soffel} 
M. Soffel, B. M\"uller and W. Greiner, \emph{Phys. Rev. D} \textbf{22} (1980), 1935 - 1937.
\bibitem{Hughes}
R. J. Hughes, {\it Annals. Phys.} {\bf 162} (1985), 1 - 30.
\bibitem{CaDe}
P. Candelas and D. Deutsch, {\it Proc. R. Soc. London Ser. A.} {\bf 362} (1978), 251 - 262. 
\bibitem{IyKu}
B. R. Iyer and A. Kumar, {\it J. Phys. A.} {\bf 13} (1980), 469 - 478.
\bibitem{Hacyan}
S. Hacyan, {\it Phys. Rev. D.} {\bf 33(12)} (1986), 3630 - 3633.
\bibitem{Takagi}
S. Takagi, Prog. Theo. Phys. Supplement No. 86, 1986.
\bibitem{BD}
N.D. Birrell and P.C.W. Davies, {\it Quantum Fields in Curved Spacetime} (Cambridge University Press, 1982).
\bibitem{Horibe}
M. Horibe, {\it Prog. Theor. Phys.} {\bf 61(2)} (1979), 661 - 671.
\bibitem{SrPa}
L. Sriramkumar and T. Padmanabhan, {\it Class. Quantum Grav.} {\bf 13} (1996), 2061 - 2079.
\bibitem{Sanchez81A}
N. Sanchez, {\it Phys. Lett. B.} {\bf 105(5)} (1981), 375 - 380.
\bibitem{Sanchez79}
N. Sanchez, {\it Phys. Lett. B.} {\bf 87(3)} (1979), 212 - 214.
\bibitem{Sanchez81B}
N. Sanchez, {\it Phys. Rev. D.} {\bf 24(8)} (1981), 2100 - 2110.
\bibitem{Yang}
Zhu Jian-yang, Bao Aidong and Zhao Zheng, \emph{Intl. J. Theoret. Phys.} \textbf{34}(10) (1995), 2049 - 2059.
\bibitem{SaWh}
N. Sanchez and B. F. Whiting, {\it Phys. Rev. D.} {\bf 34(4)}, (1986), 1056 - 1071.
\bibitem{Dolby2}
C. E. Dolby, `A State-Space Based Approach to Quantum Field Theory in Classical Background Fields', Thesis, Geometric Algebra Research Group, Cambridge University.
\bibitem{Dolbyradar}
C. E. Dolby and S. F. Gull, \emph{Am. J. Phys.} \textbf{69}, (2001), 1257 - 1261.
\bibitem{Dolby}
C. E. Dolby and S. F. Gull, \emph{Annals. Phys.} \textbf{293} (2001), 189 - 214.
\bibitem{Misner2}
C. W. Misner, K. S. Thorne and J. A. Wheeler, {\it Gravitation} (Freeman 1973). 
\bibitem{Pauri}
M. Pauri and M. Vallisneri, \emph{Found. Phys. Lett.} \textbf{13}(5) (2000), 401 - 425.
\bibitem{Bondi}
H. Bondi, {\it Assumption and Myth in Physical Theory} (Cambridge University Press, 1967).
\bibitem{Bohm}
D. Bohm, {\it The Special Theory of Relativity} (W. A. Benjamin, 1965).
\bibitem{Dinverno}
R. D'Inverno, {\it Introducing Einstein's Relativity} (Oxford University Press, 1992).
\bibitem{Dolby3}
C. E. Dolby and S. F. Gull, `Radar Time and a State-Space Based Approach to Quantum Field Theory in Gravitational and Electromagnetic Backgrounds', preprint number gr-qc/0207046.
\bibitem{QEDSF}
W. Greiner, B. M\"uller and J. Rafelski, {\it Quantum Electrodynamics of Strong Fields, with an Introduction to Modern Relativistic Quantum Mechanics}, section 21 (Springer-Verlag, 1985).
\end{thebibliography}
\end{document}